\documentclass[final,5p]{elsarticle}

\usepackage{lineno,hyperref}
\modulolinenumbers[5]

\journal{Physics Letters A}

\let\today\relax
\makeatletter
\def\ps@pprintTitle{%
    \let\@oddhead\@empty
    \let\@evenhead\@empty
    \def\@oddfoot{\footnotesize\itshape
         {Submitted preprint} \hfill\today}%
    \let\@evenfoot\@oddfoot
    }
\makeatother

\usepackage{graphicx}
\usepackage{bm,bbm}
\usepackage[dvipsnames]{xcolor}

\usepackage{amsmath}
\usepackage{amsfonts}
\usepackage{amssymb}
\usepackage{epstopdf}

\newcommand{\One}{\mathbbm{1}}
\newcommand{\rmd}{\mathrm{d}}
\newcommand{\rme}{\mathrm{e}}
\newcommand{\rmi}{\mathrm{i}}

\newcommand{\la}{\langle}
\newcommand{\ra}{\rangle}

\begin{document}
\begin{frontmatter}

\title{Quantum transport in a combined kicked rotor and quantum walk system}
\author[GDL]{Adrian Ortega}
\cortext[mycorrespondingauthor]{Corresponding author}
\ead{adrian.ortegar@alumnos.udg.mx}

\author[GDL]{Thomas Gorin}
\author[Praga]{Craig S. Hamilton}

\address[GDL]{Departamento de F\'isica, Universidad de Guadalajara, Blvd. Gral. Marcelino Garc\'ia Barrag\'an 1421, C.P. 44430, Guadalajara, Jalisco, M\'exico}
\address[Praga]{FNSPE, Czech Technical University in Prague, Br\^{e}hov\'a 7, 119 15, Praha 1, Czech Republic}

\begin{abstract}
We present a theoretical and numerical study of the competition between two
opposite interference effects, namely interference-induced ballistic transport
on one hand, and strong (Anderson) localization on the other. While the
former effect allows for resistance free transport, the latter brings the
transport to a complete halt. As a model system, we consider the quantum kicked 
rotor, where strong localization is observed in the discrete 
momentum coordinate. In this model, we introduce the ballistic transport in the 
form of a Hadamard quantum walk in that momentum coordinate.
The two transport mechanisms are combined by alternating the 
corresponding Floquet operators.

Extending the corresponding calculation for the kicked rotor, we estimate the
classical diffusion coefficient for the combined dynamics. Another argument,
based on the introduction of an effective Heisenberg time should then allow to estimate the localization time and the localization length. While this is
known to work reasonably well in the kicked rotor case, we find that it fails
in our case. While the combined dynamics still shows localization, it takes 
place at much larger times and shows much larger localization lengths than 
predicted.

Finally, we combine the kicked rotor with other types of quantum walks, namely
diffusive and localizing quantum walks. In the diffusive case, the localizing 
dynamics of the kicked rotor is completely canceled and we get pure diffusion.
In the case of the localizing quantum walk, the combined system remains 
localized, but with a larger localization length.

\end{abstract}


\end{frontmatter}

\section{\label{I} Introduction}

Interference effects at the border between classical and quantum transport
have been of particular interest as they are capable of changing the
transport properties of a system completely. One classic example is the strong 
localization (Anderson localization)~\cite{And58}, where the destructive 
interference between many random paths leads to a complete halt of transport. 
Modifications of strong localization have been studied in the last couple of 
years, for instance, the inclusion of nonlinear effects in the Anderson
model~\cite{GarciaMataPRE2009} and weak nonlinearity combined with a static
field~\cite{GarciaMataSNP2009}. The inclusion of such effects weakens the
strong localization.

Another, more recent example, is that of quantum (random) 
walks~\cite{Kem03,portugal2013quantum} which were first been proposed in
Ref.~\cite{AhDaZa93}. There, the interference between several paths may lead to
quite the opposite effect, changing the transport from diffusive to ballistic.
Applications for this effect can be found 
in the area of quantum computing and quantum versions of classical Monte Carlo 
search algorithms~\cite{portugal2013quantum} (and references therein). \\

\noindent
In this paper, we study the competition of two wave phenomena, which, from a 
quantum transport perspective, lead eventually to opposite results. On the one 
hand, ``strong localization'', which essentially brings quantum transport to a 
halt, and on the other hand the Hadamard quantum walk 
(QW)~\cite{portugal2013quantum}, which may turn a classically diffusive process
into one of ballistic transport. A convenient arena for this competition is the 
quantum kicked rotor (KR)~\cite{CasChi79,Izra90}, which shows strong 
localization in the discrete momentum coordinate, and where different types of
quantum walk can be implemented in a natural way.  

Both, the KR and the QW can be realized experimentally on a number of 
experimental platforms~\cite{ManaiPRL2015,Hai18,RyanPRA2005,NitscheNJP2016}.
In addition, recently, it has been shown that the QW dynamics can be realized
in a KR system, when the kick period is in resonance with the period of the
rotor~\cite{GroiseauJPA2018, DadrasPRL2018,DadrasPRA2019}. \\



\noindent
In the case of the KR, strong localization is not due to disorder but to 
quantum chaos. Consequently, one rather speaks about ``dynamical
localization''~\cite{CasChi79,ChiShe86,Izra90}. Still, it is possible to map
the dynamics on a tight binding model with quasi-random disorder, similar to
the one-dimensional Anderson model~\cite{GrPrFi84,Reichl2004}. 


A quantum walk (QW) may be considered as the quantum version of a classical 
random walk~\cite{Kem03,portugal2013quantum}. While it can be discrete or 
continuous in time, we limit ourselves to the discrete case. 
A classical random walker on a one-dimensional lattice, chooses at each step
to move either to the left or to the right. If this choice is random, the 
corresponding dynamics are diffusive and the variance of the walkers position increase linearly with time. 
In contrast, the quantum
walker chooses between stepping to the left or stepping to the right, depending
on the state of a two-level quantum system -- the quantum ``coin''. By making
sure that this coin is always in a superposition state, the quantum walker 
will perform a superposition of both steps. In this case the variance of the 
walker position increases quadratically in time, which is known as ballistic 
motion. If we perform measurements upon the walker, or if there is a general 
decoherence process, such as measurements of the walkers position, then the 
quantum walk ``collapses'' to the classical random walk, and diffusive motion 
is recovered~\cite{Kendon2007,Weiss15}. 


In both cases, quantum KR and QW, the dynamics is generated by the repeated
application of discrete (Floquet) evolution operators, $F_{\rm kr}$ and 
$F_{\rm qw}$, respectively. Therefore, we generate the dynamics of the 
combined system by simply alternating the two Floquet operators. The QW is 
implemented in the discrete momentum coordinate, since it is there where the 
dynamical localization of the quantum KR takes place. To vary the relative 
strength between the two models, we use the kick strength in the KR, which 
controls the localization length of the dynamics. In addition, we consider 
different step sizes for the QW.

Additional control of the dynamics of the combined system is achieved by 
choosing the operations on the quantum coin (a two level system which controls
the direction of the quantum walker) in different ways: (i) the so called 
``Hadamard'' QW, where the coin operation is a Hadamard gate, and the resulting
QW is ballistic; (ii) coin operations which are random in space and time, so
that the resulting QW is diffusive; (iii) coin operations which are random in
space only, so that the resulting QW is localized. Mainly, we focus on the
standard Hadamard QW case, however, in a final section, we also
analyse the case where the QW part is either diffusive or localized. \\

\noindent
The paper is organized as follows. After the introduction (Sec.~\ref{I}), we
review the localization properties of the quantum KR in Sec.~\ref{K}, and 
discuss the discrete time quantum walk (QW) in Sec.~\ref{W}. Sec.~\ref{M} 
contains 
a detailed analysis (numerical and analytical) of the localization properties 
of the combined system. We end the paper in Sec.~\ref{C} with our conclusions.

\section{\label{K} Kicked rotor}

Here, we review the diffusion and localization properties of the kicked rotor.
In Sec.~\ref{KC} we discuss the classical kicked rotor, in Sec.~\ref{KQ} the
corresponding quantum system. The classical kicked rotor can be reduced to a 
one-parameter family of dynamical maps, while the quantum kicked rotor has an
additional independent parameter which is reminiscent to Planck's constant.

\subsection{\label{KC} Classical kicked rotor}

For future reference, and for an unambiguous definition of the variables and 
parameters to be used throughout this work, we shortly review the dynamical
equations which define the kicked rotor.
Its dynamics are described in two-dimensional phase space, which consists of 
the angle variable, $-\pi \le \theta < \pi$, and the angular momentum
variable, $L\in\mathbb{R}$.  The Hamiltonian reads
\begin{equation}
	H(L,\theta)= \frac{L^2}{2I} + K\; \cos\theta\; 
   \sum_{n\in\mathbb{Z}} \delta(t_{\rm ph} - n\, T) \; ,
\end{equation}
where $K$ is the kick strength, $T$ the kick period and $t_{\rm ph}$ the time
in physical units. In the absence of kicks, $L$ is constant and 
$\theta(t) = \theta(0) + L t_{\rm ph}/I$. 
That means that between two kicks, $\theta$ changes as follows:
\begin{equation} 
\theta' = \theta + \frac{L\, t_{\rm ph}}{I} + 2\pi m\; , 
\end{equation}
where $m\in\mathbb{Z}$ is chosen such that $\theta$ remains inside the
interval $[-\pi, \pi)$. The kick by contrast is instantaneous and leaves 
$\theta$ unchanged. However, for $L$ we find
\begin{equation}
L' = L + K\; \sin\theta \; . 
\end{equation}
Putting both processes together (starting with the kick), we find
\begin{equation}
  \begin{split}
  L_{t+1} &= L_t + K\; \sin\theta\; , \\
  \theta_{t+1} &= \theta_t + \frac{T}{I}\; L_{t+1}
  = \theta_t + \frac{T}{I}\; \big (\,  L_t + K\; \sin\theta\, \big )\; ,
\end{split}
\end{equation}
where the integer $t$ measures time in units of the kick period $T$.
Multiplying the first equation with $T/I$, and redefining $p_t = T\, L_t/I$,
we find
\begin{equation}
\begin{split}
   p_{t+1} &= p_t + \kappa\; \sin\theta \; ,\\
   \theta_{t+1} &= \theta_t + p_t + \kappa\; \sin\theta\; , 
\end{split}
\label{KC:krmap}\end{equation}
where $\kappa = TK/I$. This shows, that the classical kicked rotor
is essentially a one-parameter family of dynamical maps, parametrized by 
$\kappa$, see e.g.~\cite{Delande2013}.

\begin{figure}
\includegraphics[width= 0.5\textwidth]{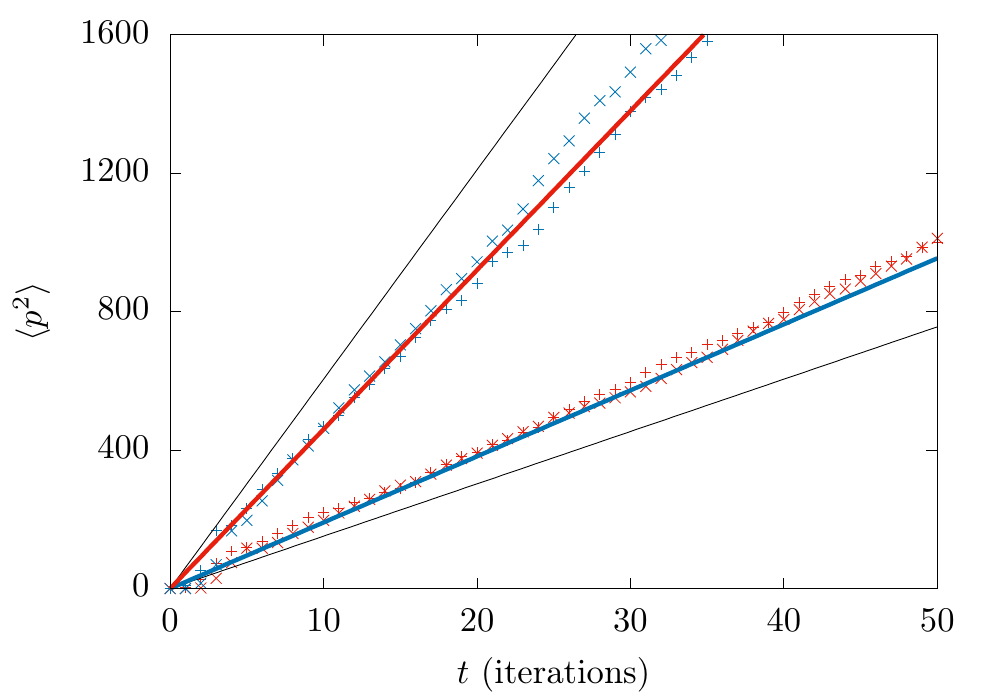}
\caption{Second moment of the momentum as a function of time, showing classical
diffusion, for $\kappa = 5.5$ (red points) and $11.0$ (blue points). The solid 
black lines show the simple classical expectation according to 
Eq.~(\ref{KC:theodiffus}), the colored straight lines are obtained from the 
improved theoretical estimate for the diffusion constant in 
Eq.~(\ref{KC:theodiffus2}) -- for $\kappa = 5.5$ (blue line) and 
$\kappa = 11.0$ (red line). We averaged over $N_{\rm sam} = 1000$ trajectories 
with initial conditions in phase space, chosen from a central square region of 
size $0.1\times 0.1$ ($+$ symbols) and $1.0\times 1.0$ ($\times$ symbols), 
respectively. }
\label{f:ClassDiffus}\end{figure}

\paragraph{Diffusion in momentum space} We compute trajectories for the kicked
rotor map given in Eq.~(\ref{KC:krmap}) for $\kappa = 5.5$ and $\kappa = 11.0$ 
and plot the average kinetic energy as a function of time (number of 
iterations). As initial conditions, we choose initial points uniformly in a 
square region of side lengths $\Delta$ around the center of phase space.
A simple argument which assumes statistical independence of subsequent 
iterations yields:
\begin{equation}
\la p^2_{t+1}\ra = \frac{\kappa^2}{2}\; \la p^2_t\ra \; .
\label{KC:theodiffus}\end{equation}
However, in Ref.~\cite{ChiShe86} this estimation has been replaced by the more
accurate expression
\begin{equation} 
\sigma^2_p(t) = \la p_t^2\ra - \la p_t\ra^2 \approx D_0\; t \; , 
\end{equation}
with the classical diffusion constant (in momentum space) given as
\begin{equation}
D_0 = \frac{1}{2}\begin{cases}
   \kappa^2\, \big [\, 1 - 2\, J_2(\kappa) + 2\, J_2(\kappa)^2\, \big ] &:
      \kappa \ge 4.5 \\
   0.6\; (\kappa - \kappa_{\rm cr})^3 &: \kappa_{\rm cr} < \kappa < 4.5
   \end{cases}\; .
\label{KC:theodiffus2}\end{equation}
Here, $\kappa_{\rm cr} \approx 0.9716$ and $J_2(x)$ is the Bessel
function~\cite{AbrSte70}. 

In Fig.~\ref{f:ClassDiffus}, we illustrate the diffusion in the momentum 
coordinate of the classical kicked rotor for two different values of $\kappa$.
These values, $\kappa =5.5$ ($\kappa = 11.0$) are chosen in such a way that the 
more precise estimation for the diffusion coefficient $D$ is above (below) the 
simple estimates (solid black lines) from Eq.~(\ref{KC:theodiffus}). It can be 
seen that the numerical data follow the improved expression from 
Eq.~(\ref{KC:theodiffus2}) rather accurately.

\subsection{\label{KQ} Quantum kicked rotor}

The Hamiltonian in the quantum case is given by
\begin{equation}
\frac{\hat L^2}{2I} + K \cos\hat\theta\; 
   \sum_{n\in\mathbb{Z}} \delta(t_{\rm ph} - n\, T) \; , \qquad
\hat L = -\rmi\, \hbar\, \partial_\theta \; .
\label{KQ:DefHam}\end{equation}
The time evolution between two kicks is given as
\begin{equation} 
U_{\rm free} = \rme^{-\rmi \hat L^2\, T/(2I\, \hbar)}\quad\text{with}\quad
   [\hat L, \hat\theta] = \rmi\, \hbar \; .  
\end{equation}
Let us rescale the angular momentum operator $\hat L$ and introduce a
dimensionless effective Planck constant $\hbar_{\rm eff}$. Then we may write
\begin{equation}
\hat p = \frac{T}{I}\; \hat L\; , \quad \hbar_{\rm eff} = \frac{\hbar\, T}{I}
\quad :\quad
U_{\rm free} = \rme^{-\rmi \hat p^2/(2 \hbar_{\rm eff})}\; , 
\label{commutrel}\end{equation}
with the new commutation relation 
$[\hat p, \hat\theta] = \rmi\, \hbar_{\rm eff}$. The kick itself affects the 
system via another unitary operator, namely
\begin{equation} 
U_{\rm kick} = \rme^{-\rmi K\, \cos(\hat\theta)/\hbar} 
 = \rme^{-\rmi \kappa\, \cos(\hat\theta)/\hbar_{\rm eff}}\; , 
\end{equation}
which can be seen by solving the Schr\" odinger equation in position (i.e. 
angle) representation (see~\ref{apD}), while using the dimensionless 
kick strength $\kappa$, as defined in Eq.~(\ref{KC:krmap}).

\paragraph{Momentum representation}
The eigenstates of $\hat p$ are the periodic plane waves,
\begin{equation}
    \varphi_m(\theta) = \frac{\rme^{\rmi m\, \theta}}{\sqrt{2\pi}}
\end{equation}
such that
\begin{equation}
\hat p\; \varphi_m(\theta) 
 = -\rmi\hbar_{\rm eff}\, \partial_\theta\; \varphi_m(\theta) 
 = \hbar_{\rm eff}\, m\; \varphi_m(\theta)
\end{equation}
with $m\in\mathbb{Z}$. This allows to write
\begin{equation} 
U_{\rm free} = \sum_{m\in\mathbb{Z}} |\varphi_m\ra\; 
      \rme^{-\rmi \hbar_{\rm eff}\, m^2/2}\; \la\varphi_m| \; . 
\label{KQ:UfreeOfm}\end{equation}
In what follows, we use the simpler notation $|m\ra = |\varphi_m\ra$ for the 
eigenstates of the momentum operator.

It remains to find the momentum representation of the operator $U_{\rm kick}$.
For that purpose, we need to evaluate the following integral:
\begin{equation} 
\la m| U_{\rm kick}\, |n\ra = \frac{1}{2\pi}\int_0^{2\pi}\rmd\theta\;
   \rme^{\rmi (n-m)\, \theta}\; \rme^{-\rmi\kappa/\hbar_{\rm eff}\ \cos\theta}
   \; . 
\end{equation}
This is achieved with the help of the formula~\cite{AbrSte70},
\begin{equation} 
\rme^{-\rmi z \cos\theta} 
   = \sum_{k\in\mathbb{Z}} (-\rmi)^k\; J_k(z)\; \rme^{\rmi k\theta} \; , 
\end{equation}
which yields
\begin{equation}
  \begin{split}
&\la m| U_{\rm kick}\, |n\ra = \sum_{k\in\mathbb{Z}} (-\rmi)^k\; J_k(z)\;
   \frac{1}{2\pi}\int_0^{2\pi}\rmd\theta
   \rme^{\rmi (n-m)\, \theta}\; \rme^{\rmi k\theta}\\
&\qquad = \sum_{k\in\mathbb{Z}} (-\rmi)^k\; J_k(z)\; \delta_{n-m+k} 
 = (-\rmi)^{m-n}\; J_{m-n}(z) \;,
  \end{split}
\label{KQ:UkickBessel}\end{equation}
where we have set $z =\kappa/\hbar_{\rm eff}$. In our simulations of
the quantum kicked rotor, we use the symmetrized version of 
the Floquet operator~\cite{Izra90}, 
\begin{equation}
F_{\rm kr} = U_{\rm free}^{1/2}\; U_{\rm kick}\; U_{\rm free}^{1/2} \; ,
\label{KQ:symFkr}\end{equation}
and for the evolution of quantum states $\Psi(t)$, we use
\begin{equation}
\Psi(t+1) = F_{\rm kr}\; \Psi(t)\; ,
\end{equation}
where the integer $t$ measures time in units of the kick period $T$.

\subsection{\label{KL} Localization}

Localization, or more precisely, Anderson localization~\cite{And58}, is the
absence of wave diffusion in a disordered medium. The effect is usually
related to the fact that the eigenstates of the system are exponentially 
localized in space. This gives rise to the definition of the ``localization
length'' in terms of the average exponential envelope of the eigenstates.

The quantum KR shows this type of localization in the momentum coordinate.
In this case, the effect is called ``dynamical localization''. While the
classical KR shows normal diffusion in the momentum coordinate (e.g. a linear
increase of the energy with time), cf. Eq.~(\ref{KC:theodiffus2}), this 
diffusion breaks down in the quantum case. 
The effect as such has been observed first in Ref.~\cite{CasChi79}. It has been
explained with the Anderson localization in Ref.~\cite{ChiShe86}; see also 
Ref.~\cite{FiGrPr82,GrPrFi84}.

\subsubsection*{Semi-quantitative theoretical description} 

The analytical estimation of the localization length of the KR is based on the
following heuristic argument, which consists of two 
steps~\cite{CIS81,ChiShe86,CIS88, Izra90,Delande2013,ManRob13}:
(i) estimation of the localization time $t_{\rm loc}$, i.e. the time when 
localization set in, and (ii) calculation of the ``shape'' of an evolving 
quantum state in the localized regime, i.e. at times larger than the 
localization time. 

\paragraph{(i) Estimation of the localization time}
The basic idea is to identify the localization time $t_{\rm loc}$ with the 
(effective) Heisenberg $t_{\rm H}$ time of the system, i.e. the time where the 
system ``realizes'' that the spectrum is discrete. For time-periodic systems 
this time is given as $t_{\rm H} = 2\pi/d$, where $d$ is the average spacing 
between the complex eigenvalues of the Floquet operator~\cite{Haake2001}. 

One may then argue that $t_{\rm loc} = t_{\rm H}$ is the time when localization 
sets in. In a simplified picture, the quantum mechanical momentum uncertainty 
increases until $t \sim t_{\rm loc}$, where the momentum uncertainty freezes 
due to localization. Thus, 
\begin{align}
\sigma^2_p 
  &= \la\Psi(t)| \hat p^2\, \Psi(t)\ra - \la\Psi(t)|\hat p\, \Psi(t)\ra^2
\notag\\
  &\approx D_0\; \min(t,t_{\rm loc})\; ,
\label{KL:diffloctrans}\end{align}
where $\Psi(t)$ is the evolving quantum state which is typically taken as 
starting out from a momentum eigenstate (be reminded that $t$ measures time in
units of the kick period $T$ and is therefore discrete).

Unfortunately, it is impossible to estimate the Heisenberg time
directly, because $F_{\rm kr}$ as given in 
Eq.~(\ref{KQ:symFkr}), has a dense spectrum. This problem is circumvented by
considering only the ``relevant'' eigenstates for the evolution of a given 
initial state. 

Due to the expected exponential localization of these eigenstates, the relevant
eigenstates must be sufficiently close to the momentum $p_0$ of the initial 
state. In other words, these eigenstates must be localized in the momentum 
interval
$(p_0 - \hbar_{\rm eff}\, l_\infty \, ,\, p_0 + \hbar_{\rm eff}\, l_\infty)$.
Then, the Weyl law (or EBK quantization) yields~\cite{Haake2001}
\begin{equation} 
\mathcal{N} \approx \frac{2\pi\, 2\hbar_{\rm eff}\, l_\infty}
      {2\pi\, \hbar_{\rm eff}} = 2\, l_\infty 
\end{equation}
as the approximate number of relevant eigenstates for the evolution of the 
system. Therefore, the nearest neighbor distance between the relevant 
eigenvalues of the Floquet operator may be approximated as 
$d \approx 2\pi/\mathcal{N}$, which yields 
\begin{equation}
t_{\rm loc} = \frac{2\pi}{d} \approx 2\; l_\infty \; .
\label{tlocresult}\end{equation}
Note that this result is based on an ad hoc numerical lower limit for the 
overlap between the relevant eigenstates and the initial state. Changing this 
numerical limit will change $t_{\rm loc}$ accordingly. 

\paragraph{(ii) Shape and momentum variance of the evolving quantum state}
At times larger than the localization time, one expects that the momentum 
uncertainty for the quantum state $\Psi(t)$ remains approximately constant. The 
envelope of the state (in the momentum representation) should remain constant 
also, only the individual coefficients remain fluctuating in time. In
Refs.~\cite{CIS81,ChiShe86,CIS88}, the following expression has been derived:
\begin{equation}
\big | \la m|\Psi(t)\ra \big |^2 \approx \frac{1 + 2|m - \bar m|/l_s}{2 l_s}
  \; \rme^{-2|m - \bar m|/l_s} \; ,
\label{KQ:Psitmdist}\end{equation}
where $l_s \approx 2\, l_\infty$. This formula may be interpreted as a 
probability distribution, which is symmetric with respect to its center, 
$\bar m$. In the case that $\Psi(t)$ starts out from a momentum eigenstate with 
quantum number $m_0$, we expect that $\bar m$ is close to $m_0$. The 
distribution is normalized such that 
\begin{align}
\sum_m \big | \la m|\Psi(t)\ra \big |^2 &\approx \int_{-\infty}^\infty\rmd x\;
   \frac{1+2x}{2}\; \rme^{-2|x|} = 1 \; , \notag\\
\sum_m (m - m_0)^2\; \big | \la m|\Psi(t)\ra \big |^2 &\approx 
   l_s^2\int_{-\infty}^\infty \rmd x\; x^2\; \frac{1+2x}{2}\; \rme^{-2|x|} 
\notag\\
 &= l_s^2 \; .
\end{align}
Using Eq.~(\ref{KL:diffloctrans}) allows us to express $\sigma^2_{p, \rm loc}$,
the momentum variance of the evolving state after localization, in terms of 
$l_\infty$:
\begin{equation}
\sigma^2_{p, \rm loc} \approx \hbar_{\rm eff}^2\; l_s^2 
  \approx D_0\; 2\, l_\infty
  \quad :\quad  l_s \approx \frac{D_0}{\hbar_{\rm eff}^2}\; .
  \label{eq:lsD}
\end{equation}
With this result, we can calculate the remaining unknown quantities:
\begin{equation}
t_{\rm loc} \approx l_s \approx \frac{D_0}{\hbar_{\rm eff}^2} \; , \qquad
\sigma^2_{p , \rm loc} \approx D_0\, l_s 
   \approx \frac{D_0^2}{\hbar_{\rm eff}^2} \; .
\label{KQ:tloc}\end{equation}
Unless stated otherwise, we will consider the case $\hbar_{\rm eff} = 1$, in
the remainder of the paper.

\subsubsection*{Numerical results}

In what follows, we show simulations for the quantum KR, where we evolve 
momentum eigenstates in time. We study the typical shape of these states 
far in the localized regime (at time $t= 10\, t_{\rm loc}$), and the behavior 
of the average momentum variance $\overline{\sigma_p^2}$ as a function of time.
Unless stated otherwise, we consider a sample of $103$ different initial 
states with momentas taken from the range $m_0 \in [680,782]$.

\begin{figure*}
  \includegraphics[width=0.9\textwidth]{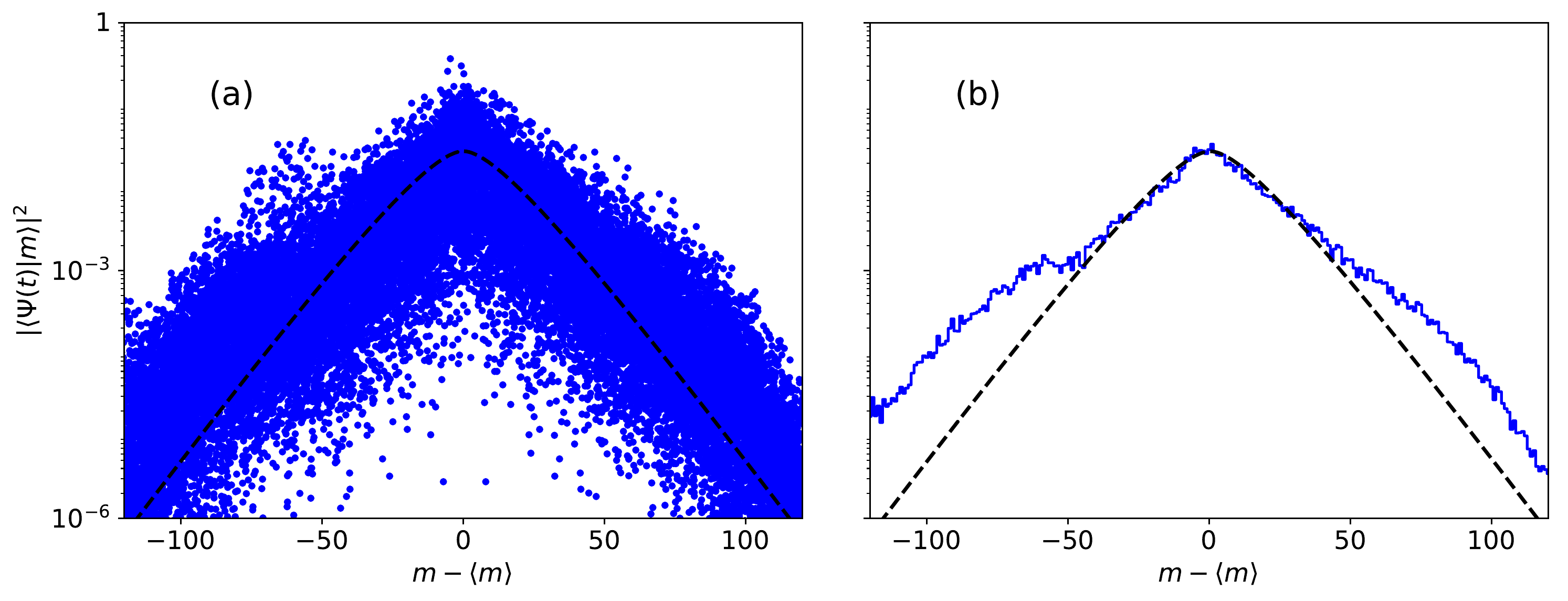}
\caption{Average shape of momentum eigenstates evolved up to 
  $10\times t_{\rm loc}$, for $\kappa = 5.5$ in a logarithmic scale. 
  For details about re-centering and averaging, see text.
  Panel (a): individual squared expansion coefficients, 
  $|\la m|\Psi_{m_0}(t)\ra|^2$ for $m_0 \in [680,782]$ (blue points). 
  Panel (b): Histogram generated from the data shown in panel (a), by defining
  bins of size one and accumulating the squared expansion coefficients to yield
  an average probability distribution, which can be compared to 
  Eq.~(\ref{KQ:Psitmdist}) (blue line). In both panels, the dashed black line
  shows the best fit to Eq.~(\ref{KQ:Psitmdist}), with $l_s = 18.0$. }
\label{f:KQ:fig1}\end{figure*}

\paragraph{Wavefunction shapes}
Numerically, we consider the time evolution of $|\Psi(t)\ra$ for the sample of 
initial states defined above. To calculate the average shape of these 
states we plot the probabilities $|\la m|\Psi(t)\ra|^2$ versus the relative 
momentum coordinate, $m' = m - \la m\ra$, where 
\begin{equation}
\la m\ra = \sum_m m\; |\la m|\Psi(t)\ra|^2
         = \frac{1}{\hbar_{\rm eff}}\; \la\Psi(t)| \hat p\, \Psi(t)\ra\; . 
\end{equation}
We then convert the data into a histogram, defining bins of unit length around 
the integer values of $m'$, summing up all probabilities which fall into the 
respective bin, normalizing the resulting histogram at the end. 

In Fig.~\ref{f:KQ:fig1} we show the result of this procedure for $\kappa = 5.5$
and $t = 10\, t_{\rm loc}$ on a logarithmic scale. For each member in the 
ensemble, we obtain $|\langle \Psi(t)|m\rangle|^2$ as a function of the 
(centered) momentum $m'$. The results for all members in the ensemble are shown 
in Panel (a). Panel (b) shows the 
corresponding accumulated histogram (blue line) together with the best fit to
Eq.~(\ref{KQ:Psitmdist}) (black dashed line). We find that this function 
agrees well with the numerical histogram in the center of the distribution. In
the tails we find larger deviations, however these only have little
statistical weight.

\begin{figure*}
 \includegraphics[width=0.9\textwidth]{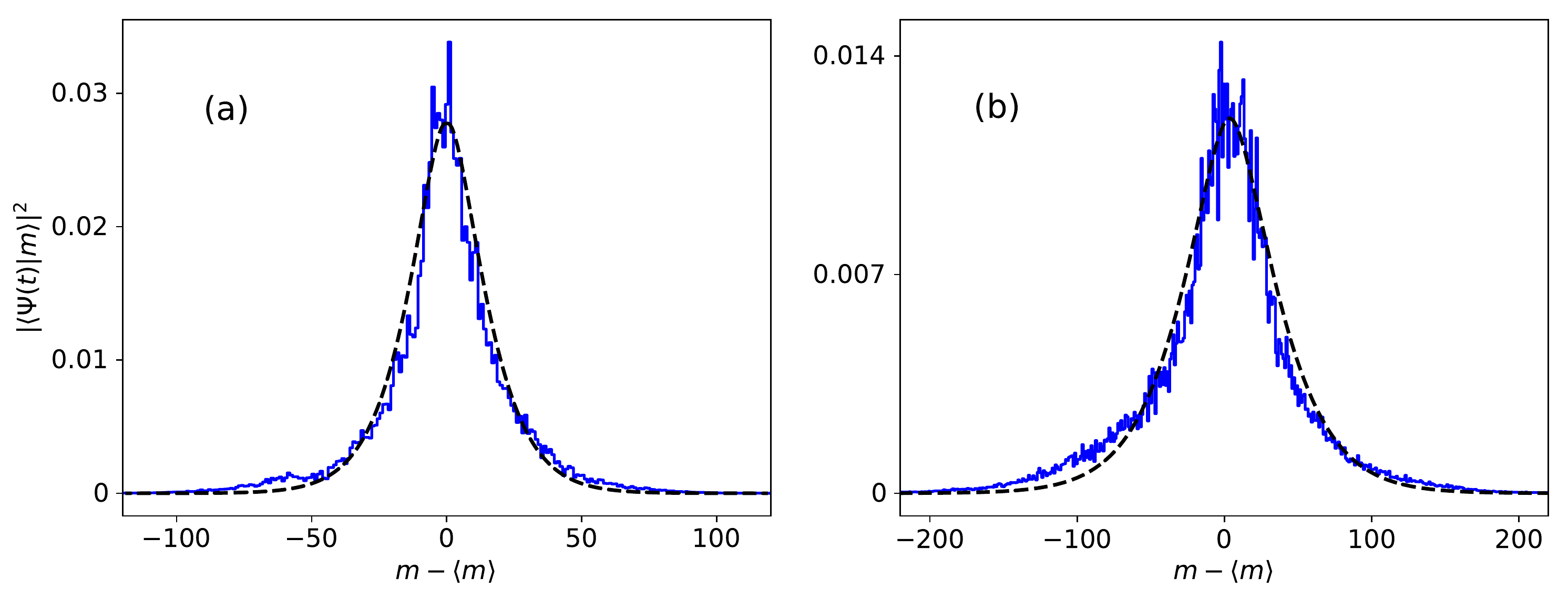}
 \caption{Comparison of the average shapes of momentum eigenstates evolved up
   to $10\times t_{\rm H}$, for $\kappa = 5.5$ (left panel) and 
  $\kappa = 11.6$ (right panel). 
   Here, we use a linear scale. The set of the initial states and the basis
   size are the same as in Fig.~\ref{f:KQ:fig1}. The histograms (blue lines)
   and the fit function (dashed black line) are computed as in Fig.~\ref{f:KQ:fig1}:   $l_s = 18.0$ for $\kappa = 5.5$ and $l_s = 41.7$ for $\kappa = 11.6$.}
\label{f:KQ:fig2}\end{figure*}

In Fig.~\ref{f:KQ:fig2} we compare the average shapes of momentum eigenstates,
evolved up to $t= 10\, t_{\rm loc}$, for $\kappa = 5.5$ (a) and
$\kappa = 11.6$ (b). The dashed curves correspond to best fitting theoretical
estimate of Eq.~\ref{KQ:Psitmdist}. For $\kappa=5.5$ we find that the best
fitting parameter, i.e. the localization length, is $l_s=18$, while for
$\kappa=11.6$, $l_s=41.7$. There are some differences between the theoretical
estimate and the actual profiles, as already pointed out in~\cite{Izra90}.

\begin{figure*}
  \includegraphics[width=0.495\textwidth]{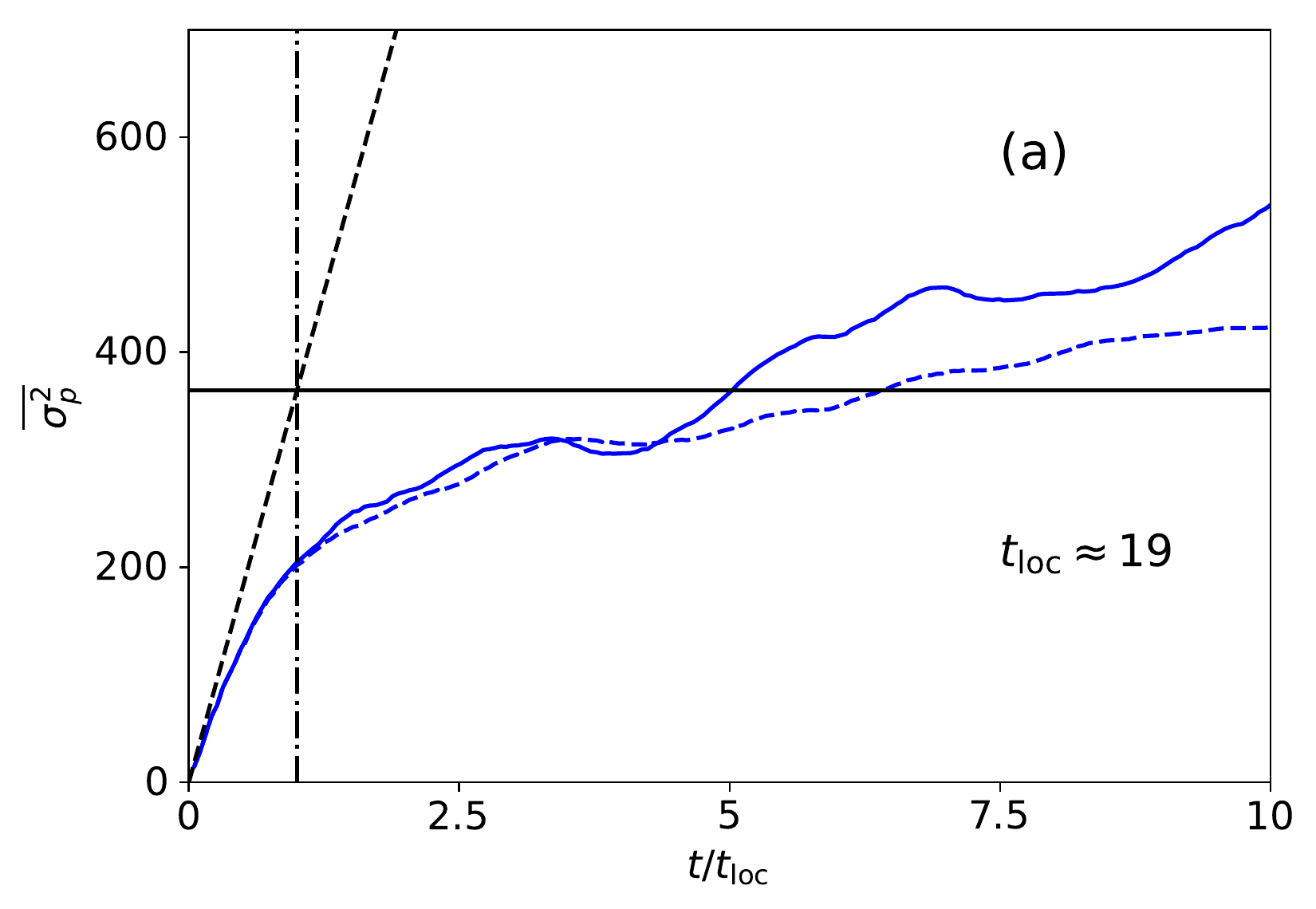}
  \includegraphics[width=0.495\textwidth]{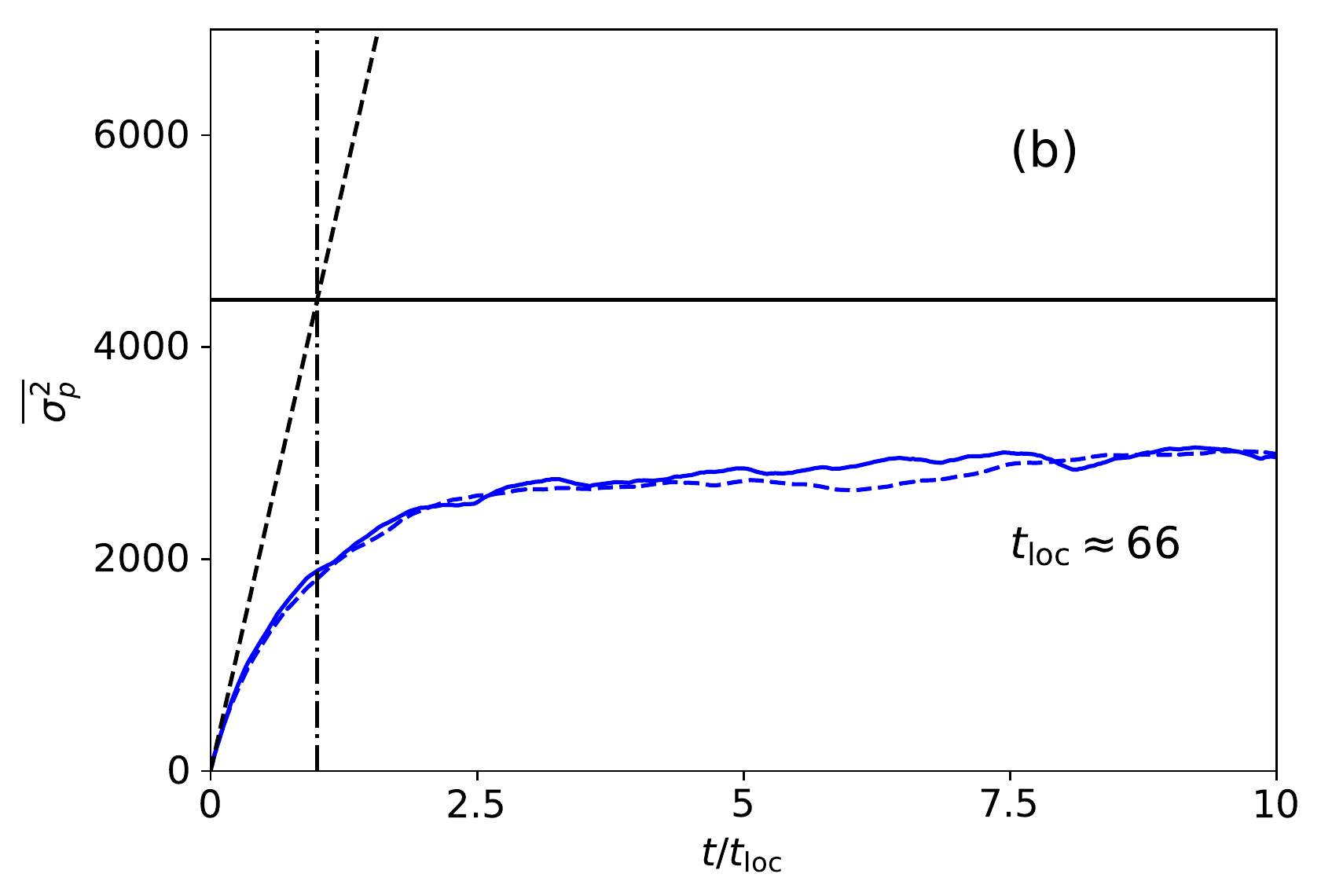}
\caption{Ensemble averaged $\overline{\sigma^2_p}$ for the kicked rotor, as a function 
of discrete time $t$. Blue solid line for the small sample $m_0 \in [680,782]$; 
blue dashed line for the large sample $m_0 \in [800,1800]$. In panel (a) 
$\kappa=5.5$ and in panel (b) $\kappa=11.6$. 
For comparison we have included the classical difussion (black dashed line), 
the theoretical saturation of $\sigma^2_p$ (horizontal line) and $t_{\rm loc}$ 
(vertical line). }
\label{f:KQ:fig3}\end{figure*}

\paragraph{Momentum variance as a function of time}
Here, we compute the ensemble averaged time evolution of the momentum variance 
$\overline{\sigma_p^2}$. The results are shown in 
Fig.~\ref{f:KQ:fig3} (blue lines) for $\kappa = 5.5$ in panel (a) and for 
$\kappa=11.6$ in panel (b). For comparison, we indicate the key quantities,
related to the theoretical description of the localization effect: the expected 
classical diffusion according to Eq.~(\ref{KL:diffloctrans}) (black dashed 
line); the expected saturation value of the momentum variance, 
$\sigma^2_{p, \rm loc}$ as defined in Eq.~(\ref{eq:lsD}) (black solid line); 
and $t_{\rm loc}$ from Eq.~(\ref{tlocresult}) (black dash-dotted line).

In both cases, $\kappa =5.5$ [panel (a)] and $\kappa = 11.6$ [panel (b)],
the numerical results show notable deviations from the 
theoretical prediction. While the transition from the diffusive regime to 
localization is clearly happening at the expected time $t \sim t_{\rm loc}$, 
the numerical curves do not converge to the expected saturation value 
$\sigma^2_{p, \rm loc}$. On panel (a) $\kappa = 5.5$, both curves (for the 
ensembles averages over $103$ and $10^3$ states) overshoot the saturation value
quite notably. On panel (b) $\kappa = 11.6$, by contrast, the corresponding
curves remain well below. 
Even by analyzing much longer times and additional values for $\kappa$, we 
could not arrive at more consistent results. It seems that the theoretical 
model described above only provides a semi-quantitative description of 
localization in the kicked rotor.

We believe that the agreement would improve at
smaller values for $\hbar_{\rm eff}$. In essence, the theoretical model is
based on semi-classical arguments applied to a quantum-chaotic system. One
should therefore expect an improvement when choosing smaller values for
$\hbar_{\rm eff}$. Unfortunately, even reducing $\hbar_{\rm eff}$ by a
half increases the computational cost prohibitively, in particular in the 
case of the combined system KR plus quantum walk, to be discussed below.



\section{\label{W} Quantum walk}

In contrast to the kicked rotor (KR), the quantum walk (QW) dynamics have no 
``simple'' Hamiltonian description. Instead, one defines a unitary operator which
is applied at each discrete point in time. This description matches nicely with 
that of the quantum kicked rotor, where we repeatedly apply the Floquet 
operator $F_{\rm kr}$, defined in Eq.~(\ref{KQ:symFkr}). 

The QW dynamics requires an additional two-level quantum system, the quantum 
``coin'', which steers the quantum walker. In the simplest case, the full 
Hilbert space is given by the momentum coordinate and the quantum coin. 
Then, we define the unitary operation:
\begin{equation}
\begin{split}
U_{\rm qw} &= S_\rho\; (\, U_{\rm c} \otimes \One\, ) \; , \\
S_\rho &= |0\ra\la 0| \otimes \hat D_\rho 
   + |1\ra\la 1| \otimes \hat D_{-\rho} \; ,
\end{split}
\label{W:UqwSimple}\end{equation}
where $U_{\rm c}$ is a unitary operator in the coin system, $\One$ is the 
identity operation in the momentum coordinate, and $\hat D_\rho$ is the 
discrete displacement operator 
\begin{equation}
\hat D_\rho\; |m\ra = |m+\rho\ra \; , \qquad 
D_\rho^\dagger = \hat D_{-\rho}\; ,
\label{defDisplaceOp}\end{equation}
in the momentum coordinate. 
Unless stated otherwise, we will restrict ourselves to the case $\rho=1$. 
Analogous to the kicked rotor, the evolution of a quantum state from time $t$
to time $t+1$ is obtained by 
\begin{equation}
|\Psi(t+1)\rangle = U_{\rm qw}\; |\Psi(t)\rangle \; .
\label{W:QWdynSimple}\end{equation}

We will also consider cases where the unitary coin operation, $U_{\rm c}$, 
is chosen at random, at different sites (i.e. for different momentas) and/or at 
different times. 
In such a case, 
$U_{\rm c} = U_{\rm c}^{(m,t)}$ and
\begin{equation}
\begin{split}
U_{\rm qw}^{(t)} &= S_\rho\; \Big (\, \sum_m U_{\rm c}^{(m,t)}\otimes
   |m\ra\la m|\, \Big )\; , \\
|\Psi(t+1)\rangle &= U_{\rm qw}^{(t)}\; |\Psi(t)\rangle \; .
\end{split}
\label{W:UqwGeneral}\end{equation}

\paragraph{Ballistic quantum walk}
One of the cases we will consider in more detail is the Hadamard quantum 
walk, where the coin operator is
\begin{equation} 
U_{\rm c} = H = \frac{1}{\sqrt{2}}
   \begin{pmatrix} 1 & 1\\ 1 & -1\end{pmatrix}\; . 
\label{defHgate}\end{equation}
This choice leads to ballistic transport, which means that 
$\la\hat p\ra$ or $\sigma_p$ increase linearly in time.%
\footnote{It is possible to choose initial states for the quantum coin, such 
that the probability distribution in momentum space, propagates mainly in only 
one direction. Then it is $\la\hat p\ra$ rather than $\sigma_p$ which increases 
linearly in time.}
 
For our simulations we choose the initial states as
\begin{equation} 
\Psi(0) = \frac{1}{\sqrt{2}}\; \begin{pmatrix} 1\\ \rmi\end{pmatrix}
   \otimes |m_0\ra\; , 
\end{equation}
that is a symmetric eigenstate of $\sigma_y\otimes\One$. This choice leads to 
two wave packets which move symmetrically and ballistically away from the 
initial site $|m_0\ra$.

\paragraph{Diffusive quantum walk}
The simplest way to obtain diffusive dynamics (where $\sigma^2_p$ increases 
linearly in time) 
consists in choosing $U_{\rm c}$ differently and at random at each time step.
To this end, we choose $U_{\rm c}$
from the invariant distribution on SU$(2)$, which means that the corresponding
probability measure is the normalized Haar measure of the group~\cite{Haar33}.

In practice, we generate random elements of this distribution by diagonalizing
random Hermitian two-by-two matrices which are chosen from the Gaussian unitary 
ensemble (GUE)~\cite{Haake2001}. If $U$ diagonalizes such a GUE matrix, i.e.
\[ H\; U = U\; \begin{pmatrix} \lambda_1 & 0\\ 0 & \lambda_2\end{pmatrix}\; ,
\]
then we choose two random phases $\theta_1$ and $\theta_2$ from the uniform
distribution on the interval $[0,2\pi)$ and set
\[ U_{\rm c} = U\; \begin{pmatrix} \rme^{\rmi\theta_1} & 0\\ 0 &  
      \rme^{\rmi\theta_1} \end{pmatrix} \; . \]
In this way we obtain a sequence of identically and independently distributed
unitary matrices $\{\, U_{\rm c}^{(t)} = U_{\rm c}^{(m,t)}\, \}$ which are used
to construct the single-step quantum-walk operators, defined in 
Eq.~(\ref{W:UqwGeneral}). Since the unitary matrices are random only in
time, the evolution of the system can be calculated as  
\begin{equation} 
U_{\rm qw}^{(t)} = S_k\; \big (\, U_{\rm c}^{(t)}\otimes\One\, \big ) \; ,
\quad \Psi(t+1) = U_{\rm qw}^{(t)}\; \Psi(t) \; .
\end{equation}

\paragraph{Disordered quantum walk with localization}
It is also possible to observe localization in the quantum walk dynamics. For
that purpose, one should associate a different coin to different sites (here
momentum eigenstates). Hence, we generate a sample of random unitary matrices
$\{\, U_{\rm c}^{(m)} = U_{\rm c}^{(m,t)}\, \}$ and construct a single-step 
quantum-walk operator. Eq.~(\ref{W:UqwGeneral}) then simplifies to
\begin{equation}
  \begin{split}
U_{\rm qw} &= S_k\; \Big (\, \sum_m U_{\rm c}^{(m)}\otimes |m\ra\la m|\, 
   \Big )\; ,\\
   \Psi(t+1) &= U_{\rm qw}\; \Psi(t) \; .
  \end{split}
\label{CoinOpDisorder}\end{equation}

\paragraph{Disorder in space and time} 
Finally, we may consider the case, where we perform random and independent
unitary coin transformations at different sites and different times. In this 
case, we generate a sample of identically and independently distributed coin
transformations $\{\, U_{\rm c}^{(m,n)}\, \}$, and use Eq.~(\ref{W:UqwGeneral})
to compute the dynamics of the system. In this case, we expect to obtain the
same type of diffusive dynamics as in the diffusive quantum walk case discussed previously.  

\begin{figure*}
  \includegraphics[width=1\textwidth]{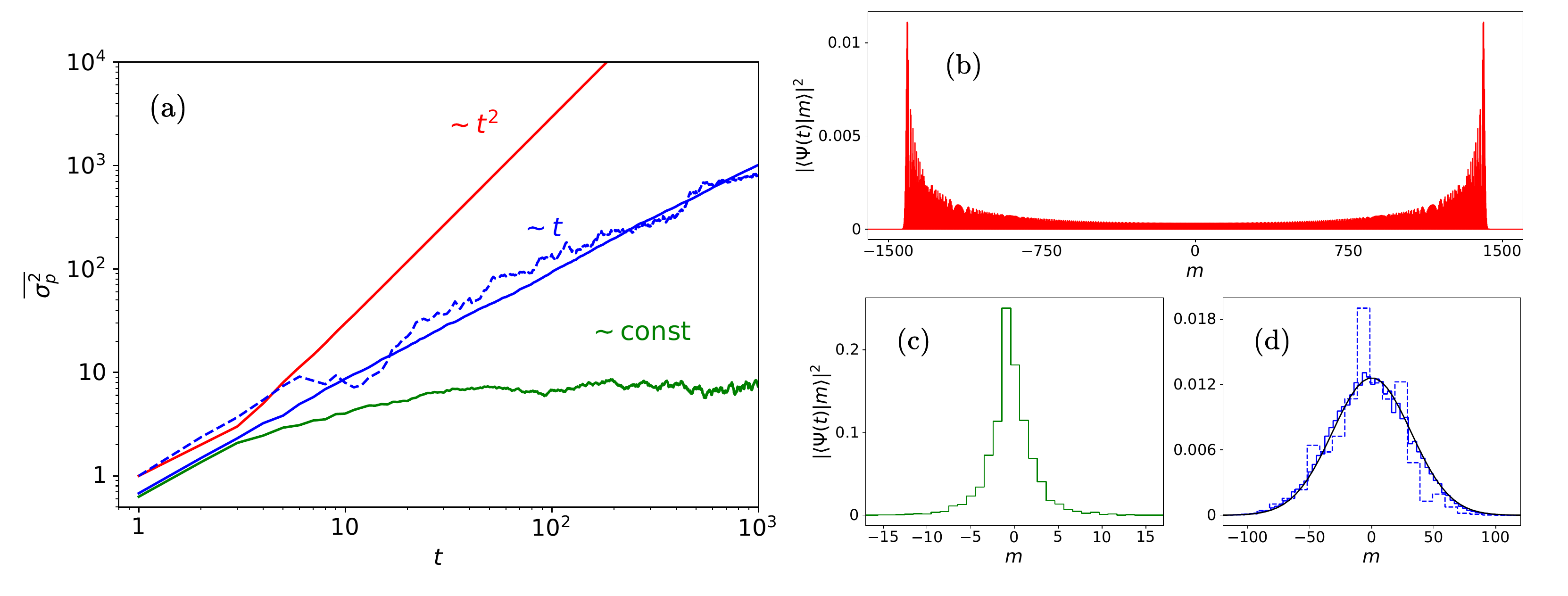}
  \caption{Panel (a): Ensemble averaged $\overline{\sigma_p^2}$ as a function 
  of the number of time steps $t$ in a double-log plot. We depict the following
  cases: ballistic Hadamard quantum walk (red line), diffusive QW (blue line)
  and QW with disorder (green line). The dashed blue line shows 
  site-independent random operations in time. Panels (b), (c) and (d): Final
  probability distributions for the respective cases shown in panel (a) with 
  the same color code. In (d) we have included the analytical
  gaussian curve in black line with variance $\sigma^2_p=D_\rho t$ with
  $\rho=1$, see Eq.~\ref{eq:diffrho}. As expected, the dynamics is equal to the classical random walk.}
\label{f:W:sig2}\end{figure*}

In Fig.~\ref{f:W:sig2}, panel (a), we show the variance $\overline{\sigma_p^2}$
(as introduced in Fig.~\ref{f:KQ:fig3})
as a function of discrete time, just as in the kicked rotor. The figure shows
the result for all 
four types of dynamics: (i) the ballistic dynamics (red line), (ii) the 
diffusive dynamics with randomness in (momentum) space and time (blue solid 
line), (iii) the diffusive dynamics with randomness in time, only (blue dashed
line), and (iv) the localized dynamics with randomness in space, only (green 
line). In the rest of the panels we plot the evolved state profiles after
$10^3$ time steps for the Hadamard walk (b), the quantum walk with onsite
disorder (c), and the diffusive dynamics with randomness in space and
time and time only in solid and dashed lines, respectively. As expected, this
reproduces earlier results e.g. those in~\cite{Schreiber2011},~\cite{brun2003}
for coined quantum walks and in \cite{Weiss15} for the KR.

\section{\label{M} Combination of kicked rotor and quantum walk}

In this section we study the competition between the two 
respective interference effects, strong localization present in
the KR and ballistic transport in the
case of the Hadamard QW (Sec.~\ref{MH}). We will then consider
cases where the QW part is either diffusive or localizing (Sec.~\ref{MR}). In
all cases, we combine the KR with the QW dynamics by alternatingly applying the 
single step unitary evolution from one model and the other:
\begin{equation}
F_{\rm qw,kr} = U_{\rm qw}\; U_{\rm free}^{1/2}\; U_{\rm kick}\; 
U_{\rm free}^{1/2}\; .
\label{eq:Ffull}
\end{equation}
Note that the operators $U_{\rm free}$ and $U_{\rm kick}$ have to be
extended by adding the identity in the Hilbert space of the two-level quantum 
coin. In addition, in the case of the diffusive quantum walk, $U_{\rm qw}$ is 
really time dependent as it contains different random coin operations at 
different times.

\subsection{\label{MH} Combination of KR and Hadamard QW}

Here, we study the dynamics generated by the Floquet operator, defined in
Eq.~(\ref{eq:Ffull}), where $U_{\rm qw}$ is constructed with the Hadamard gate
from Eq.~(\ref{defHgate}) as coin operation, and conditional displacements by 
$\rho \ge 1$ steps, as defined in Eq.~(\ref{defDisplaceOp}). 

Naturally, we expect that the ballistic quantum walk steps will counteract the 
localization of the KR. However, it remains to be seen whether the localization will just be delayed and weakened (larger localization time and larger localization
length) or whether it will be canceled; in the latter case the dynamics would
become diffusive or ballistic. In order to arrive at a more quantitative 
expectation, we compute the effect of the QW part on the classical diffusion
constant, as given in Eq.~(\ref{KC:theodiffus2}). This calculation which 
follows exactly the original procedure for the KR, is detailed 
in~\ref{appb:diff}. It yields the following result:
%
\begin{equation}
  D_\rho = \rho^2+\frac{\kappa^2}{2}\big [\, 1-2J_2(\kappa)+J_2(\kappa)^2
     \, \big ]\; ,
\label{eq:diffrho}\end{equation}
where we have assumed that $\kappa > 4.5$. Hence the only change as compared to
the pure KR expression consists in the addition of $\rho^2$. Note that for 
$\kappa > 4.5$, the KR diffusion constant is of the order of $20$ and larger.
Therefore, we would expect that a unit-step quantum walk ($\rho = 1$) will have
only a small effect on the KR dynamics.

In Sec.~\ref{KL} we describe the argument which leads to an estimate of the 
localization length for the KR, based on the classical diffusion constant 
$D_0$. With the classical diffusion constant $D_\rho$ for the combined system
at hand, Eq.~(\ref{eq:diffrho}), we follow the argument step by step and
thereby obtain a similar estimate for the localization length of the combined
system. The only problematic point is the shape of the evolving quantum states,
as it is given in Eq.~(\ref{KQ:Psitmdist}). However, as long as $\rho$ is not 
very large, it seems reasonable to assume that this shape remains approximately 
the same. We find that the result for the KR, Eq.~(\ref{KQ:tloc}), remains 
valid, if we simply replace the KR diffusion constant $D_0$ by the KR plus QW 
diffusion constant $D_\rho$. 

\begin{figure}
  \centering
  \includegraphics[width=0.48\textwidth]{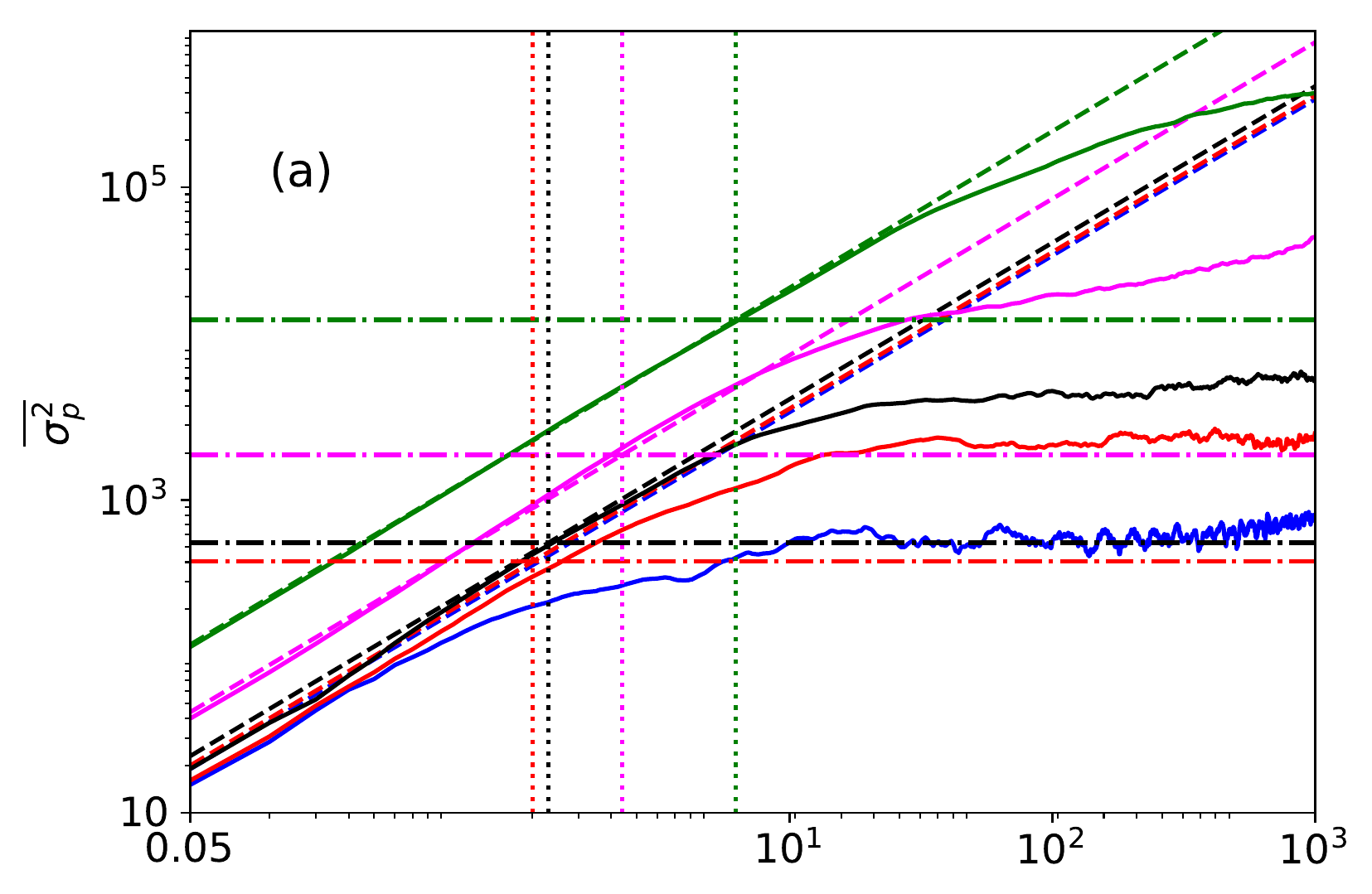}
  \includegraphics[width=0.48\textwidth]{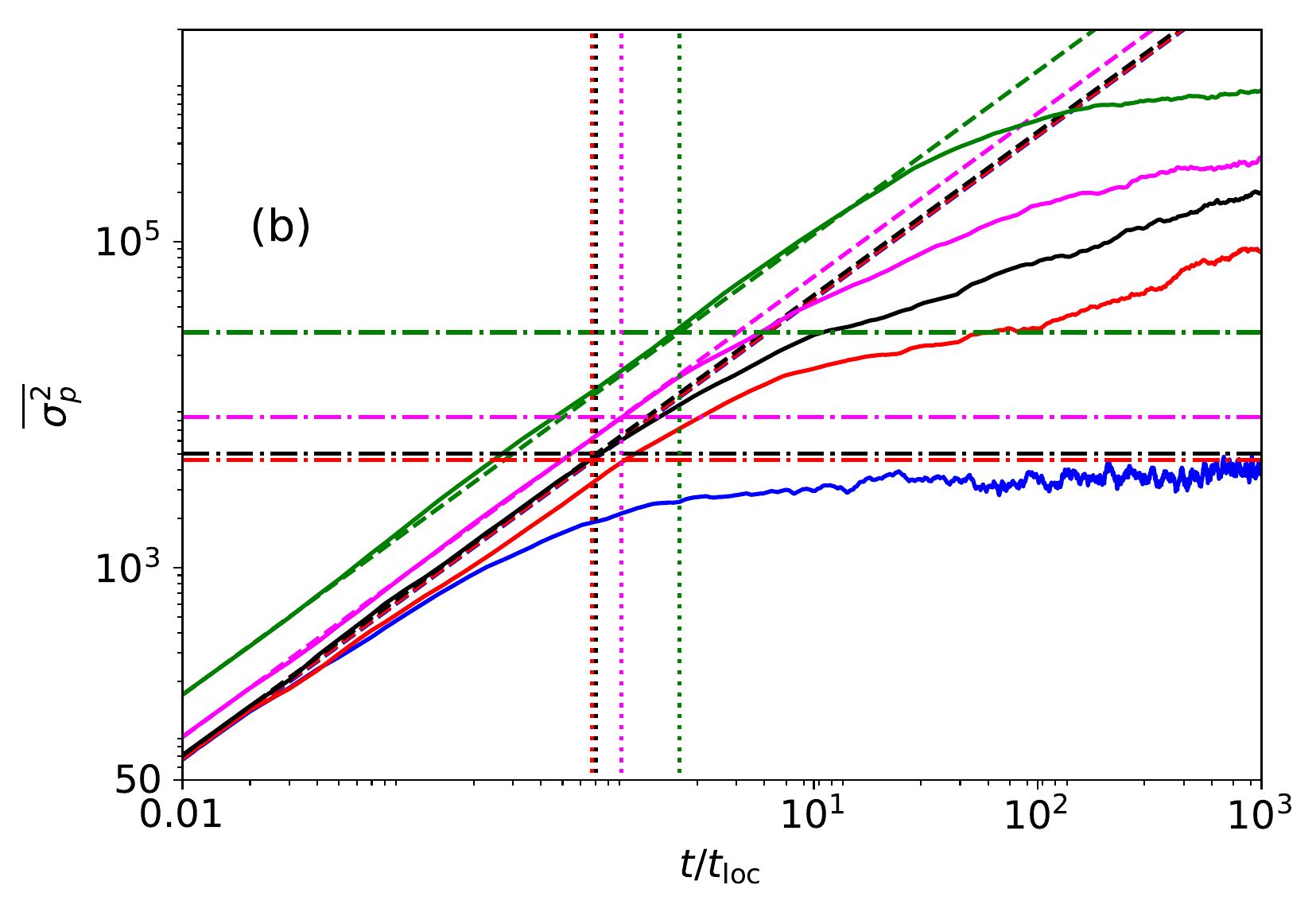}
  \caption{Variance $\overline{\sigma^2_p}$ as a function of discrete time $t$,
    for the kicked rotor plus Hadamard quantum walk. 
    In panel (a), $\kappa=5.5$ and in panel (b), $\kappa=11.6$. For comparison,
    we have included the blue solid
    line corresponds to $\rho=0$, i.e. the quantum kicked rotor. The red solid
    line corresponds to $\rho=1$, black solid line to $\rho=2$, pink solid
    line $\rho=5$ and green solid line $\rho=10$. The dashed lines correspond
    to classical diffusion using the diffusion constant given in
    Eq.~\ref{eq:diffrho}. The horizontal lines corresponds to the saturation 
    value $\sigma_{p, {\rm loc}}^2$ and the vertical lines to $t_{\rm loc}$.}
  \label{f:7a}\end{figure}

We numerically analyse the localization properties of the combined system in
Fig.~\ref{f:7a}. Here, we plot the average momentum variance 
$\overline{\sigma^2_p}$ as a function of time $t$ (measured in units of the 
kick period) on a double-logarithmic scale. This allows us to cover a much larger
range in time and momentum variance as compared to Fig.~\ref{f:KQ:fig3} (where 
we studied the pure KR case). Here, panel (a) shows the case $\kappa=5.5$ and
panel (b) the case $\kappa=11.6$. The solid lines represent the
numerical evolution of the full system for $\rho=1$ (red), $\rho=2$ (black),
$\rho=5$ (pink) and $\rho=10$ (green). For completeness we have included
the kicked rotor case case $\rho=0$ in blue. The dashed lines correspond to
classical
diffusion using the diffusion coefficient $D_\rho$ as given in 
Eq~(\ref{eq:diffrho}). The horizontal and vertical lines, in the stated color
code, corresponds to the saturation variance $\sigma_p^2$ and $t_{\rm loc}$,
respectively.

In all cases, the numerical results for the average momentum variance initially shows 
 the expected diffusive behaviour, where the slope is in reasonable 
agreement with the diffusion constant $D_\rho$, obtained in 
Eq~(\ref{eq:diffrho}). At larger times, eventually all numerical curves deviate
from the straight line, which indicates the transition to the localizing 
regime. However, even though we extend the simulations to very large times, 
namely $10^3$ times the KR localization time $t_{\rm loc}$, only for 
$\kappa = 5.5$ and $\rho =1$ (and may be $\rho = 2$ we find a clear tendency 
of the average momentum variance to saturate, and in all cases the expected
theoretical saturation values are exceeded by at least an order of magnitude.

\subsection{\label{MR} Combination of KR and different random QWs}

Here, we study the dynamics when the QW part is constructed using random
coin operations. We distinguish two cases, the ``diffusive QW'' where the 
coin operations are chosen at random in (momentum) space and time, see 
Eq.~(\ref{W:UqwGeneral}), and the ``disordered QW'', where the coin operations 
are chosen at random in (momentum) space, only, see Eq.~(\ref{CoinOpDisorder}). 
In this part, we limit ourselves to unit-step displacements, $\rho = 1$.

\begin{figure}
\centering
\includegraphics[width=0.5\textwidth]{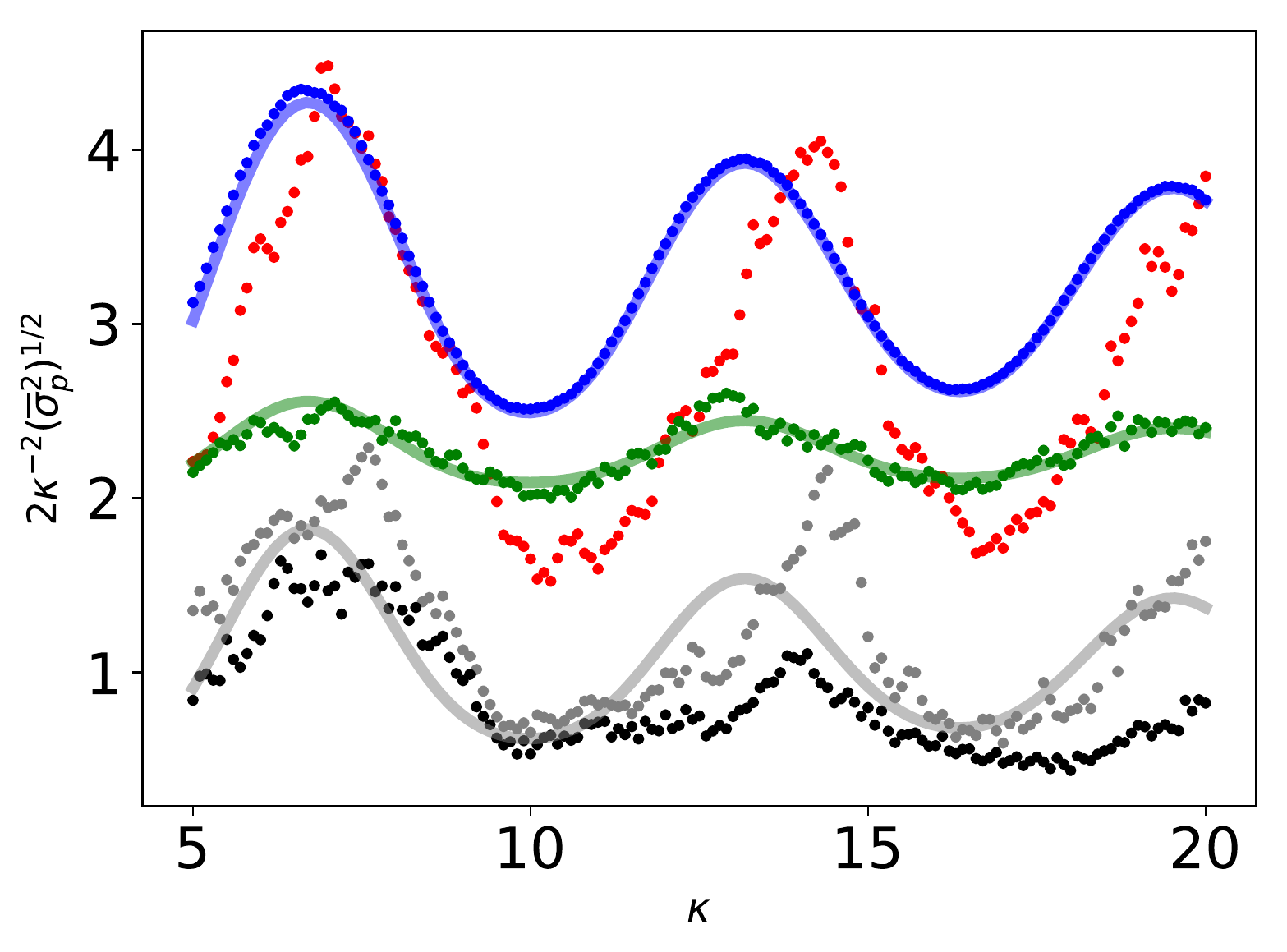}
\caption{ Scaled standard deviation of the momentum operator for 
the combination of the KR with different types of QW as a function of the kick
strength $\kappa$.
For details of its computation and the extra factor $(\kappa^2/2)^{-1}$ see the 
main text. KR only (gray points), KR plus Hadamard QW (red dots), KR plus
diffusive QW (blue dots), KR plus QW with on-site disorder (green dots).
The black dots correspond to the quantity $2\kappa^{-2}l_s$, where $l_s$ is 
obtained with the help of Eq.~(\ref{eq:lsD}), as in Fig.~\ref{f:KQ:fig2}.
Where available, we included theoretical estimates (solid lines of the same
color as the corresponding numerical results); for details see the main text. }
\label{f:7}\end{figure}

To quantify the effect of the different types of QWs on the KR dynamics, we 
consider the square root of $\bar\sigma^2_p$ at the time $t= 10\, t_{\rm loc}$ 
as a function of $\kappa$. Based on Eq.~(\ref{KQ:tloc}) we expect for the pure 
KR dynamics that 
\begin{equation}
\overline{\sigma^2_p} \approx D_0^2\; ,
\end{equation}
while we expect deviations for the combined system, KR plus QW.
To show these deviations on a convenient scale, we follow
Ref.~\cite{ChiShe86} and multiply $(\, \overline{\sigma^2_p}\, )^{1/2}$ by 
$2/\kappa^2$, which takes away the leading trend of the $\kappa$-dependence 
(in the KR case). The result is plotted in Fig.~\ref{f:7}. Note 
that in all cases shown, the curves seem to oscillate around an average 
constant value. This means that the overall dependence on $\kappa$ remains the 
same as in the pure KR case. Below, we discuss the results for each case 
individually.
\begin{itemize}
\item[(i)]
The pure KR case is shown with gray dots. This can be compared to the solid 
gray line which shows the theoretical expectation based on the knowledge of the 
diffusion constant, as it is given in Eq.~(\ref{KC:theodiffus2}). The agreement 
is rather semi-quantitative. This is not surprising in view of 
the deviations found already in Fig.~\ref{f:KQ:fig3}.
\item[(ii)]
For the KR plus Hadamard QW case, discussed in the previous Sec.~\ref{MH}, the 
result is shown with red points. Note that there is no obvious reason why the 
theoretical considerations which lead us to Eq.~(\ref{KQ:tloc}) could not be 
applied analogously to the present case, as well. Therefore,
one expects the effect of the QW part onto the KR dynamics to be rather small, 
due to the small change in the corresponding diffusion constant (note that
$\rho = 1$). In fact the theoretical curve, based on Eq.~(\ref{eq:diffrho}) 
would be hardly distinguishable from the solid gray curve, in particular at
large values of $\kappa$. In spite of this, the numerical results show very
strong differences, with pronounced anharmonic oscillations and on average much
larger values.
\item[(iii)]
The green points 
KR plus QW with on-site disorder as defined in Eq.~(\ref{CoinOpDisorder}). 
In this case, the standard deviation of the momentum seems to be approximately 
equal to 
\begin{equation}
\sqrt{\bar\sigma^2_p}\; \frac{2}{\kappa^2} \approx 2 \quad :\quad
\sqrt{\bar\sigma^2_p} \approx \kappa^2 \; ,
\end{equation}
plus some weakened remanent of the characteristic oscillations from the KR. As
a guide to the eye, the solid green line shows the function 
$a + b\; [1 -2 J_2(\kappa) + J_2(\kappa)^2]$, with best fit values $a=1.85$ and 
$b=0.387$. Also here we see a strong effect, even though we just noted that
in classical terms, the quantum walk part is only a small perturbation.
This result also shows that combining two different mechanisms for
localization does not yield stronger, but weaker localization. Indeed, for the
pure localizing QW, we have $\overline{\sigma_p^2}$, which implies a very small
localization length as compared to the kicked rotor case.
Nevertheless the localization length of the combined system is clearly larger
than in the pure KR case, for some values of $\kappa$ more than twice as large.

\item[(iv)]
Finally, the blue points show the case where the QW part is diffusive, i.e. 
the coin operation is random in (momentum) space and time. Here one expects 
that the randomness in time destroys all phase coherences between different 
paths, and thereby cancels strong localization. 
In this case, we find that the result is consistent with the assumption of
normal diffusion with a diffusion coefficient equal to $D' \approx \kappa^2/2$.
This can be seen from the fact that at time $t= 10\, t_{\rm loc}$ we find:
\begin{align} 
\overline{\sigma^2_p} &= 10\, t_{\rm loc}\; D' = 10\; D_0\; D'\; , \notag\\
\frac{2}{\kappa^2}\; \big (\, \overline{\sigma^2_p}\, \big )^{1/2} 
  &\approx \sqrt{10\, D_0}\; \frac{\sqrt{2}}{\kappa} \notag\\
  &\approx \sqrt{10\, \big [\, 1 - 2\, J_2(\kappa) + 2\, J_2(\kappa)^2\, 
     \big ]}\; .
\label{diffusivcaseFig7}\end{align}
In Fig.~\ref{f:7} this function is shown as a solid blue line, which agrees 
very nicely with the corresponding numerical results (blue points). 
\end{itemize}

\section{\label{C} Conclusions}

In this work, we studied the competition between two fundamentally different 
interference effects, namely strong localization on the one hand and 
interference induced ballistic transport on the other. For that purpose, we 
choose the quantum kicked rotor (KR) as the base system, and add discrete-time 
quantum walk (QW) steps. The implementation of the QW steps requires an 
additional two-level system, the so called ``quantum coin''. In experimental 
cold-atom realizations of the kicked rotor, internal atomic states may be used
for that purpose. At the end, the system evolves by alternatingly applying the 
Floquet operator of the KR and the QW step (except for the diffusive case, the
QW step is just another Floquet operator).

In a preliminary part, we review in some detail the localization properties of
the KR. This study reveals some rather unexpected deviations from the
theoretical expectation, namely in the case of average wavefunction shapes and 
the corresponding localization lengths. In the main part of the paper, we then 
analyse the effect of adding QW steps to the KR dynamics.

We find that only the diffusive QW is able to destroy the localization 
completely. The ballistic and localizing QW steps increase the localization
length and as a consequence also the saturation value of the momentum variance.
Using a convenient rescaling, we analyze the average momentum variance 
$\overline{\sigma_p^2}$ as a function of the kick-strength $\kappa$. In the 
pure KR case, $\overline{\sigma_p^2}$ shows an overall trend which is 
proportional to $\kappa^2$ and an additional modulation which can be described
in terms of the Bessel function $J_2(\kappa)$. When adding ballistic QW steps, 
mostly the modulations are increased, by contrast, when adding localizing QW 
steps, the overall trend is increased while the modulations are damped. As a 
consequence, there are regions on the $\kappa$ axis, where the localization 
length is larger when the QW steps are ballistic and others (smaller ones) 
where it is larger when the additional QW steps are localizing.

Finally, we adapt the analytical calculation of the classical diffusion
coefficient for the KR to include the ballistic QW time steps.
The result shows that typically, the additional QW steps constitute a small
perturbation (in classical terms) to the KR dynamics, which modify the
diffusion coefficient only a ittle. This remains true for diffusive and localizing QW steps, also. In quantum mechanical terms, however, the QW steps
constitute a very strong perturbation, and indeed lead to strong effects on the
transport properties of the system.
The adapted analytical calculation of the classical diffusion coefficient, can
be used as input to the semiclassical argument which is commonly used to obtain
an analytical prediction of the localization length in the KR case. However,
while the argument works reasonably well for the KR, we find that it fails by
orders of magnitude (c.f. Fig.~\ref{f:7a}) when applied to the combined system.
A careful analysis of the semiclassical argument applied to the 
combined KR+QW system might shed new light on its limits of validity and even 
point at new options for improving its accuracy. 

\section*{Acknowledgements}

We gratefully acknowledge Ignacio Garcia-Mata, Sonja Barkhofen and
Andreas Buchleitner for fruitful discussions.
C.~S.~H. received support from the Grant Agency of the Czech Republic under 
Grant No. GACR 17-00844S, Ministry of Education RVO 68407700 and Centre for 
Advanced Applied Sciences, Registry No.~CZ.02.1.01/0.0/0.0/16\_019/ 0000778, 
supported by the Operational Programme Research, Development and Education, 
co-financed by the European Structural and Investment Funds and the state 
budget of the Czech Republic.

\begin{appendix}
\section{\label{apD} Discrete Fourier transform}

In principle, our Hilbert space is that of $2\pi$-periodic square integrable
functions, with the scalar product
\begin{equation}
\la\psi|\phi\ra = \int_0^{2\pi}\rmd\theta\; \psi(\theta)^*\; \phi(\theta) \; .
\end{equation}
Let us now choose an integer $N \gg 1$, fixed but arbitrary, and replace the
exact scalar product by the following discretized version:
\begin{equation}
\la\psi|\phi\ra = \frac{2\pi}{N}\sum_{k=0}^{N-1} \psi(\theta_k)^*\; 
   \phi(\theta_k) \; , \qquad \theta_k = 2\pi\; \frac{k}{N} \; ,
\end{equation}
where the prefactor $2\pi/N$ comes from the discretization of the diferential,
$\rmd\theta$. Then, for sufficiently well behaved (e.g. piecewise continous
and square integrable) functions, the discretized scalar product converges to
the original one, for sufficiently large $N$.

With this, we can build a new approximate Hilbert space which consists of 
complex finite sequences, $(\psi_k)_{0\le k < N}$ (which approximate the 
original wave functions), and the discretized scalar product we had just
introduced:
\begin{equation}
\la\psi|\phi\ra = \frac{2\pi}{N}\sum_{k=0}^{N-1} \psi_k^*\; \phi_k \; .
\end{equation}
In this new $N$-dimensional Hilbert space, consider the following $N$ state
vectors:
\begin{equation}
|\varphi_m\ra = \big (\, \varphi^{(m)}_k\, \big )_{0 \le k < N} \; , \quad
   \varphi^{(m)}_k= \frac{1}{\sqrt{2\pi}}\; \rme^{2\pi\rmi\, km/N} \; ,
\label{KF:momestates}\end{equation}
for $0\le m < N$. Note that on the one hand, these states are approximations to
the angular momentum eigenstates in the original continous Hilbert space, as
\begin{equation}
\varphi^{(m)}_k = \varphi_m(\theta_k) = \frac{1}{\sqrt{2\pi}}\; 
   \rme^{\rmi m \theta_k} \; , \qquad \theta_k = 2\pi\; \frac{k}{N} \; .
\end{equation}
But on the other hand they are also an exact ortho-normal basis in our 
approximate $N$-dimensional Hilbert space. This is because
\begin{equation}
\la\varphi_m|\varphi_n\ra = \frac{2\pi}{N}\sum_{k=0}^{N-1} \frac{1}{2\pi}\;
   \rme^{2\pi\rmi\, k (n-m)/N} = \delta_{nm} \; .
\end{equation}

\paragraph{Momentum representation} 
In the approximate Hilbert space, we started with the position (angle) 
representation,
\begin{equation}
|\psi\ra \;\;\to\;\; (\psi_k)_{0\le k < N} \; , \qquad \psi_k = \psi(\theta_k)
\; , \quad \theta_k = 2\pi\; \frac{k}{N} \; .
\end{equation}
However, we can now use the new ortho-normal basis in order to obtain an 
alternative representation:
\begin{equation}
  \begin{split}
|\psi\ra &= \sum_{m=0}^N c_m\; |\varphi_m\ra \; , \\
c_m &= \la\varphi_m|\psi\ra 
   = \frac{2\pi}{N}\sum_{k=0}^{N-1} \frac{1}{\sqrt{2\pi}}\; 
   \rme^{-2\pi\rmi km/N}\; \psi_k \; .
  \end{split}
\end{equation}
This is precisely the discrete Fourier transform. We define this as the forward
Fourier transform, which converts the position (angle) representation of a 
quantum state into its momentum representation (as we will see below, the 
states $|\varphi_m\ra$ are eigenstates of the (angular) momentum operator.
Finally, together with the inverse Fourier transform, we have
\begin{equation}
  \begin{split}
c_m &= \frac{\sqrt{2\pi}}{N}\; \sum_{k=0}^{N-1}\; 
   \rme^{-2\pi\rmi km/N}\; \psi_k \; , \\
\psi_k &= \frac{1}{\sqrt{2\pi}}\; \sum_{m=0}^{N-1}\; 
\rme^{2\pi\rmi km/N}\; c_m \; .
  \end{split}
\end{equation}

\paragraph{Momentum operator and momentum eigenvalues} Again, to be precise, we
should call this ``angular momentum operator and angular momentum 
eigenvalues''. In the original Hilbert space of continuous functions, the 
momentum eigenstates and their eigenvalues are defined as 
\begin{equation}
  \begin{split}
|\varphi_m\ra &= \frac{1}{\sqrt{2\pi}}\; \rme^{\rmi m\, \theta} \; , \\
\tilde p\; |\varphi_m\ra &= -\rmi\hbar_{\rm eff}\, \partial_\theta\; 
|\varphi_m\ra = \hbar_{\rm eff}\, m\; |\varphi_m\ra \; , 
  \end{split}
\end{equation}
and $m\in\mathbb{Z}$. Thus, we obtain the spectral representation of the
momentum operator, as
\begin{equation}
\tilde p = \sum_{m\in\mathbb{Z}} |\varphi_m\ra\; \hbar_{\rm eff}\, m\; 
   \la\varphi_m| \; .
\end{equation}
In the finite dimensional Hilbert space this formula will be approximated by
\begin{equation}
\tilde p = \sum_{m=0}^{N-1} |\varphi_m\ra\; \hbar_{\rm eff}\, m\; 
   \la\varphi_m| \; ,
\end{equation}
with the understanding, that the momentum coordinate is now periodic just in
the same way as the position coordinate. In other words, the momentum 
eigenvalue $\hbar_{\rm eff}\, m$ is exactly the same as 
$\hbar_{\rm eff}\, (m-N)$. Note that this also implies that 
$|\varphi_m\ra = |\varphi_{m-N}\ra$, as can be verified in 
Eq.~(\ref{KF:momestates}). Therefore, in order to reproduce as far as possible
the momentum coordinate of the continous case, we define (here $N$ is assumed
to be even)
\begin{equation}
  \begin{split}
\tilde p &= \!\!\sum_{m=0}^{N/2 - 1} \!\! |\varphi_m\ra\; \hbar_{\rm eff}\, m\; 
   \la\varphi_m| + \!\! \sum_{m= N/2}^{N - 1} \!\! |\varphi_m\ra\; 
   \hbar_{\rm eff}\, (m - N)\; \la\varphi_m|\notag\\
 &= \sum_{m=0}^{N/2 - 1} |\varphi_m\ra\; \hbar_{\rm eff}\, m\; 
   \la\varphi_m| + \sum_{m= -N/2}^{-1} |\varphi_m\ra\; 
   \hbar_{\rm eff}\, m\; \la\varphi_m| \\
 &= \sum_{m=-N/2}^{N/2 - 1} |\varphi_m\ra\; \hbar_{\rm eff}\, m\; 
   \la\varphi_m|\; .
  \end{split}
\end{equation}
Similarly, for $N$ being odd, we use
\begin{equation}
  \begin{split}
\tilde p &= \!\! \sum_{m=0}^{(N-1)/2} \!\! |\varphi_m\ra\; 
   \hbar_{\rm eff}\, m\; 
   \la\varphi_m| + \!\!\!\!\!\! \sum_{m= (N+1)/2}^{N - 1} \!\!\!\! 
   |\varphi_m\ra\; 
   \hbar_{\rm eff}\, (m - N)\; \la\varphi_m|\notag\\
 &= \sum_{m=0}^{(N-1)/2} |\varphi_m\ra\; \hbar_{\rm eff}\, m\; 
   \la\varphi_m| + \sum_{m= -(N-1)/2}^{-1} |\varphi_m\ra\; 
   \hbar_{\rm eff}\, m\; \la\varphi_m| \\
 &= \sum_{m=-(N-1)/2}^{(N-1)/2} |\varphi_m\ra\; \hbar_{\rm eff}\, m\;
   \la\varphi_m|\; .
  \end{split}
\end{equation}

\section{Derivation of the classical diffusion constant for the quantum
  kicked rotor plus the Hadamard quantum walk}
\label{appb:diff}
In this appendix we derive the classical diffusion constant for the KR times 
the Hadamard QW with extended jumps. The associated unitary evolution for the
combined system is given in Eq.~\ref{eq:Ffull}. In order to derive the
diffusion constant we will follow~\cite{Rechester1981} closely. Let $P(x,p,t)$
be the probability for the classical kicked rotator to be at position $x$ with momentum $p$ at time $t$. Its time
evolution is given by 
\begin{equation}
  P(x,p,t) = P(x-p,p+\kappa \sin(x-p),t-1).
\end{equation}
To this last equation, we can add directly the $\rho$-step classical random
walk (if $\rho=1$ then we have the canonical random walk) in momentum space as
\begin{equation}
  P(x,p,t) = \frac{1}{2}\big [\, P(x,p+\rho,t-1) + P(x,p-\rho,t-1)\, \big ]
  \; .
\end{equation}
We can solve this equation by using the method in~\cite{RechesterPRA1980},
which amounts to transform $P(\cdot)$ to the characteristic function
$a(\cdot)$ using the Fourier transform
\begin{equation}
  P(x,p,t) = \sum_{m=-\infty}^\infty \int_{-\infty}^\infty \rmd k\; 
      a(m,k,t)\; \rme^{\rmi mx}\; \rme^{\rmi kp}.
\end{equation}
The characteristic function is
\begin{equation}
  \begin{split}
    a(m,k,t) &= \cos(k+m \rho)\sum_{l=-\infty}^\infty J_l(|k+m|\kappa)\times\\
    &\times a(m-l \operatorname{sign}(k),k+m,t-1),
  \end{split}
\end{equation}
while the second moment of $p$ is obtained by evaluating
\begin{equation}
  \langle p^2 \rangle_t = -\frac{\partial^2}{\partial k^2}a(m,k,t)\bigg|_{m,k=0} = D t\; .
\end{equation}
The characteristic function, after  $t$ steps, is found to be
\begin{equation}
  a(m,k,T) = \cos^t(k\rho)\; F(J,t)\; ,
\end{equation}
where
$F(\cdot)$ is the characteristic function for the kicked rotor only. Taking
the second derivative with respect to $k$ of this function yields
\begin{equation}
  a'' = G''F + 2G'F' + GF''
\end{equation}
where $G=\cos^t(k \kappa)$, $G(0)=1$, $G'(0)=0$ and $G''(0)=-\kappa^2 t$. Thus
\begin{equation}
  a'' = -\kappa^2\; t\; F(0) + F''(0),
\end{equation}
with $F(0)$ a constant that we fix to one. The final expression for the 
diffusion constant is,
\begin{equation}
  D_\rho = \rho^2 + \frac{\kappa^2}{2}\left(1-2J_2(\kappa)
  + J_2(\kappa)^2 \right)\; .
\end{equation}
Evidently, this is the same as Eq.~\ref{KC:theodiffus2} plus a contant term
proportional to the size of the step squared.
\end{appendix}

\bibliography{/home/gorin/Documentos/Bib/JabRef.bib,bibliography.bib}

\end{document}